\documentclass[pra,twocolumn,showpacs,superscriptaddress,nofootinbib]{revtex4-2}

\usepackage{graphicx}
\usepackage{bm}
\usepackage{amssymb}
\usepackage{color}
\usepackage{amsmath, mathrsfs}
\usepackage{amstext}
\usepackage{float}
\usepackage[english]{babel}

\usepackage[export]{adjustbox}
\usepackage{braket}
\usepackage{cancel}
\usepackage{soul}
\usepackage{oubraces}
\usepackage{latexsym}
\usepackage[usenames,dvipsnames]{xcolor}
\usepackage[colorlinks=true,citecolor=Blue,linkcolor=RubineRed,urlcolor=Blue]{hyperref}
\usepackage{bbold}
\usepackage{bbm}

\DeclareFontFamily{OT1}{pzc}{}
\DeclareFontShape{OT1}{pzc}{m}{it}{<-> s * [0.900] pzcmi7t}{}
\DeclareMathAlphabet{\mathpzc}{OT1}{pzc}{m}{it}

\def\dbar{\,{\mathchar'26\mkern-10mu \mathrm{d}}}

\def\dg{^\dagger}
\newcommand{\beq}{\begin{equation}}
\newcommand{\eeq}{\end{equation}}
\newcommand{\beqa}{\begin{eqnarray}}
\newcommand{\eeqa}{\end{eqnarray}}

\newcommand{\bsigma}{\bar{\sigma}}

\newcommand{\Tr}{\mathrm{Tr}}

\graphicspath{{../images/}}

\begin{document}

\title{
 First Law of Quantum Thermodynamics in a Driven Open Two-Level System}

\author{Adri\'an Juan-Delgado} 
\affiliation{Donostia International Physics Center,  E-20018 San Sebasti\'an, Spain}
\affiliation{Centro de Física de Materiales, Centro Mixto CSIC-UPV/EHU, E-20018 San Sebasti\'an, Spain}
 
\author{Aur\'elia Chenu}
\email{aurelia.chenu@uni.lu}
\affiliation{Department of Physics and Materials Science, University of Luxembourg, L-1511 Luxembourg, G.D. Luxembourg}
\affiliation{Donostia International Physics Center,  E-20018 San Sebasti\'an, Spain}
\affiliation{Ikerbasque, Basque Foundation for Science, E-48013 Bilbao, Spain}


\begin{abstract}  
Assigning the variations of internal energy into heat or work  contributions
 is a challenging task due to the fact that these properties are trajectory dependent. A number of proposals have been put forward for open quantum systems  following an arbitrary dynamics. 
 We here focus on  non-equilibrium thermodynamics of a two-level system  and explore,  in addition to the conventional approach,
 two definitions motivated by either classical work or heat,  
 in which  the driving Hamiltonian or the trajectory itself are respectively used to set up a reference basis.  
 We first give the thermodynamic properties for an arbitrary dynamics and illustrate the results on the Bloch sphere. Then, we solve the particular example of a  periodically driven qubit interacting with a dissipative and decoherence bath. 
 Our results illustrate the trajectory-dependent character of heat and work, and how contributions originally assigned to dissipation in the Lindblad equation can become coherent part assigned to work. 
 \end{abstract}
\maketitle

 Deriving the laws of thermodynamics from microscopic theory has been a long-time endeavor that has given rise to quantum thermodynamics, a blossoming field of research 
  that brings advances in   foundations of physics as well as experimental progress \cite{gemmer2009, binder2019, deffner2019}. Proposals for  quantum microengines  \cite{lloyd1997, scully2001, feldmann2003} have been experimentally implemented in technological platforms \cite{rossnagel2016, deng2018, maslennikov2019, peterson2019, vonlindenfels2019}. 

In this context, the definition of physical properties at the nanoscale such as energy, heat, and work, becomes all the more relevant. However, while the variation of internal energy is well defined from the total energy of a given system, its work and heat components are trajectory-dependent \cite{vilar2008,gelbwaser-klimovsky2017,niedenzu2019, bernardo2020}. These   thermodynamic \emph{process functions} become, in the quantum regime,   stochastic variables that cannot be described by  observable Hermitian operators \cite{talkner2007}. Heat is generally considered as being generated by irreversible processes steaming from random motion and can only be transferred when the system of interest interacts with some environment. In addition,  interactions blur the clear separation between the system and bath, making the distinction between heat and work all the more  ambiguous.

A widely-used framework to distinguish between the two contributions of internal energy change  is that put forward in the late 70's \cite{spohn1978, alicki1979}. This now `conventional' approach was derived in the weak coupling regime, assigning  changes of the Hamiltonian to work and variations in the state to heat. 
In turn, the definition of heat and work  in arbitrary open quantum dynamics  has  triggered a number of proposals.  
The two-point measurement of work in isolated systems \cite{talkner2007} has been extended to driven open systems \cite{hekking2013, salmilehto2014} including  strong coupling \cite{deffner2011} or arbitrary dynamics  \cite{campisi2009a, roncaglia2014, sampaio2018}. 
For work reservoir, measuring work stored in the reservoir by quantifying the ergotropy \cite{alicki2017} avoids violation of  the Carnot bound \cite{boukobza2013}.
Among other proposals to identify heat and work in the strong coupling regime are those using the Hamiltonian of mean force \cite{jarzynski2004, gelin2009, campisi2009, rivas2020} to describe the open system at equilibrium with the environment and obtain the system partition function, from which free energy and the system entropy follow;  semi-classical approaches \cite{bender2000, abe2011} that introduce the concept of a diagonal entropy \cite{polkovnikov2011b}; and operational approaches based on measurements  \cite{weimer2008,elouard2017a,strasberg2019, strasberg2019b}. 
Recently, a definition of heat has been proposed based on the von Neumann entropy \cite{alipour2019, ahmadi2019a} and  building on the concept of  reference trajectory \cite{girolami2019,alipour2020}. In this approach, part of what is `conventionally' (in the sense of \cite{alicki1979}) considered as heat   becomes assigned to work. 
Here, we analyze  the conventional approach \cite{alicki1979} together with the two approaches that are motivated by either work \cite{polkovnikov2011b} or  heat \cite{alipour2019, ahmadi2019a}, and where the driving Hamiltonian or the trajectory is used to set a reference basis, in a two-level system 
 undergoing an arbitrary open dynamics and illustrate  the specific example in a periodically driven open qubit.

\section{Heat and Work in a generic open two-level system} 
Let an open two-level system (TLS) follow an arbitrary trajectory  described by the reduced density matrix
\begin{equation} \label{rhoGeneral}
\rho_t = \sum_{i,j=\{e,g\}} \rho_t^{ij} \ket{i}\bra{j} = \frac{1}{2}(\mathbb{1}+ \vec{n}_t \cdot \vec{\sigma}),
\end{equation}
where $\rho_t^{ij} \equiv \bra{i}\rho_t \ket{j}$, $\vec{\sigma}\equiv(\sigma_x , \sigma_y , \sigma_z)$ are the Pauli matrices and $\vec{n}_t = (2\text{Re}(\rho_t^{eg}), -2\text{Im}(\rho_t^{eg}), \Delta_t)$ is the Bloch vector with $\Delta_t = \rho_t^{ee}-\rho_t^{gg}$ the population inversion. This trajectory can include the  TLS  interaction with an environment,  the only assumptions being it is trace preserving and continuous in time. It takes a diagonal form $\rho_t = n_{+,t} \ket{n_{+,t}}\bra{n_{+,t}} + n_{-,t} \ket{n_{-,t}}\bra{n_{-,t}}$ where the eigenvalues are given by the Bloch vector norm through $n_{\pm,t} = \frac{1}{2}(1 \pm n_t)$. The eigenstates read
\begin{subequations} \label{EigenbasisDensity}
	\begin{align}
    \ket{n_{+,t}}&=\cos\phi_{t} \ket{e}+e^{i\varphi_t}\sin\phi_{t} \ket{g}, \\
	\ket{n_{-,t}}&=-e^{-i\varphi_t}\sin\phi_{t}\ket{e}+\cos\phi_{t}\ket{g},
	\end{align}
\end{subequations} 
with $e^{-i\varphi_t}=\rho_t^{eg}/|\rho_t^{eg}|$, $\cos(2\phi_t) =\Delta_t / n_t$, and $\tan(2\phi_t)=2|\rho_t^{eg}|/\Delta_t$. The norm of the Bloch vector depends on the  population inversion and the amplitude of the coherence (defined as off-diagonal  terms in the  TLS basis), specifically 
\begin{equation} \label{normBlochVector}
n_t  \equiv |\vec{n}_t| = \sqrt{4|\rho_t^{eg}|^2 + \Delta_t^2},  
\end{equation}
and determines the state purity, 
\begin{equation} \label{purity}
\mathcal{P}_t \equiv \Tr(\rho_t^2)= (1+n_t^2)/2.
\end{equation}

We consider that the TLS is driven by the general Hamiltonian
\begin{equation}
H_t =  \vec{h}_t \cdot \vec{\sigma} = \sum_{k\in\{\pm\}} E_{k,t} \ket{E_{k,t}}\bra{E_{k,t}} ,
\end{equation}
where we omit any constant term shifting the  energy and with $\vec{h}_t\equiv (h_{x,t},h_{y,t},h_{z,t})\in \mathbb{R}^3$. The eigenenergies are $E_{\pm,t}=\pm h_t$, with $h_t \equiv |\vec{h}_t|$,  and  the eigenstates read
\begin{subequations} \label{EigenbasisHam}
	\begin{align}
	\ket{E_{+,t}}&=\cos\theta_{t} \ket{e}+e^{i\Theta_t}\sin\theta_{t} \ket{g}, \\
	\ket{E_{-,t}}&=-e^{-i\Theta_t}\sin\theta_{t}\ket{e}+\cos\theta_{t}\ket{g}.
	\end{align}
\end{subequations} 
The angles are defined from $ \bra{e}H_t\ket{g} \equiv |H_t^{eg}| e^{-i \Theta_t}$ in the  TLS, physical basis, and $\cos(2\theta_t) =h_{z,t}/ h_t$,  $\tan(2\theta_t)= |H_t^{eg}|/h_{z,t}$. We interpret this Hamiltonian as the one generating the unitary part of the dynamics and containing the Lamb-shift corrections \cite{BreuerBook}. 

\begin{figure}
\includegraphics{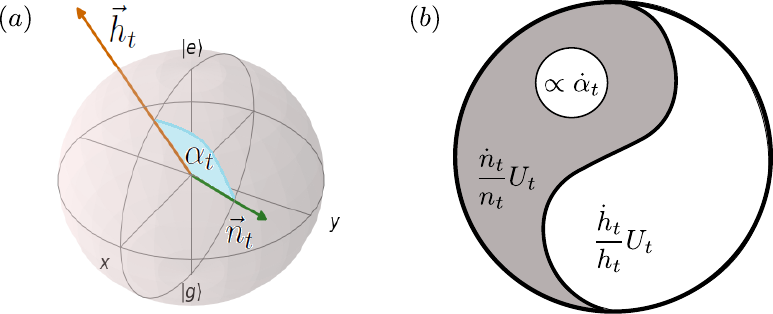}
\caption{(a) Bloch representation of the state and driving Hamiltonian. (b) Division of heat and work according the  `Hamiltonian-based' and the `entropy-based' approaches. The  former approach assigns changes of eigenenergies to work, the rest of internal energy changes being heat; in turn, the  latter approach assigns heat to changes in entropy. The difference between  these two approaches is that a portion (the white circle) of heat in the first one is attributed to work in the second one. \label{fig1} }
\end{figure}

The internal energy of the system  $U_t \equiv \Tr{(H_t \rho_t)}$ can be explicit using either the Hamiltonian basis, $\sum_k E_{k,t}  \bra{E_{k,t}} \rho_t \ket{E_{k,t}}$, or the state basis, $\sum_k n_{k,t}\bra{n_{k,t}}H_t\ket{n_{k,t}}$. It is useful to remark that it also reads  
\begin{eqnarray} \label{eq:Ut}
U_t =  \vec{n}_t \cdot \vec{h}_t  = n_t  h_t \cos\alpha_t \label{internalenergy} 
\end{eqnarray}
where $\alpha_t$ denotes the angle between  the unit vector $\hat{n}_t$ and $\hat{h}_t$, see Fig.~\ref{fig1}(a),  and we can verify that 
\begin{eqnarray}
\cos\alpha_t &&= \cos(2\phi_t) \cos(2\theta_t) +\sin(2\phi_t) \sin(2\theta_t)\cos( \varphi_t-\Theta_t ) \nonumber \\
&& = |\langle E_{+,t}\vert n_{+,t} \rangle|^2 -  |\langle E_{-,t} \vert n_{+,t} \rangle|^2. \label{eq:cosalpha}
\end{eqnarray}
While the variation of internal energy  is unambiguously defined from the time derivative of Eq. (\ref{eq:Ut}), its separation into heat and work according to the first law of thermodynamics
can be seen from different points of view and various definitions have been put forward, as mentioned in the introduction. 
 We focus on (i) today's rather conventional framework and two other approaches motivated from the classical definition of either (ii) work or (iii) heat. The first law for these three approaches reads
\begin{equation}\label{eq:firstlaw}
dU_t   = \dbar \mathbb{W}^{wc}_t + \dbar \mathbb{Q}^{wc}_t  = \dbar w_t + \dbar q_t = \dbar W_t + \dbar Q_t . \\
\end{equation}  

 (i) First, we consider the  conventional framework, that was established in the `weak-coupling' slowly-varying regime  \cite{alicki1979}. It defines work and heat  from the variation of the Hamiltonian and the trajectory,  as $\dot{\mathbb{W}}_t^{wc}=\Tr(\dot{H}_t \rho_t)$ and $\dot{\mathbb{Q}}_t^{wc}=\Tr(H_t \dot\rho_t)$, respectively. For a TLS, these read 
\begin{subequations}
\label{Alicki}
\begin{align}
\dot{\mathbb{W}}_t^{wc}&= \dot{\vec{h}}_t \cdot \vec{n}_t 
, \label{AlickiWork}\\
\dot{\mathbb{Q}}_t^{wc} &=  {\vec{h}}_t \cdot \dot{\vec{n}}_t . \label{AlickiHeat}
\end{align}
\end{subequations}
 Now, because the master equation for an open system is not unique, the dissipative part depends on the choice made for the unitary dynamics \cite{BreuerBook}. Similarly, heat and work are trajectory dependent, and contributions originally assigned to dissipation in the Lindblad equation can become coherent part assigned to work. We look at two other approaches motivated by classical thermodynamics.

(ii) Second, we consider the  `semiclassical' or `Hamiltonian-based' (HB) thermodynamics framework \cite{polkovnikov2011b}, in which  the Hamiltonian basis is  used as a reference. This approach corresponds to the classical definition of work, that  relates work to changes in the coordinates that characterize the system \cite{reichl2016}, in associating work with changes in the eigenenergies. Heat is then related to  the rest of internal energy variation, i.e. to the  variation of probabilities $p_{\pm,t} \equiv \bra{E_{\pm,t}} \rho_{t} \ket{E_{\pm,t}}=(1\pm {U}_t/h_t)/2 = (1\pm n_t \cos\alpha_t)/2$. Specifically, the changes over a small increment of time 
 read
\begin{subequations}\label{standard}
\begin{align} 
{\dot{w}}_{t} =  \sum_{k\in\{\pm\}} p_{k,t} \dot{E}_{k,t} &=\frac{\dot{h}_t}{h_t}U_t= \dot{h}_t n_t \cos \alpha_t, 
\label{standardwork} \\ 
{\dot{q}}_{t} =\sum_{k\in\{\pm\}} \dot{p}_{k,t} E_{k,t} &
= h_t \frac{d}{dt}\left( n_t \cos \alpha_t \right).    \label{standardheat} 
\end{align}
\end{subequations} 
In this framework,  entropy variations are not necessarily related to changes of heat only.  

  (iii) Third, we consider the approach where heat is defined from the change of the von Neumann entropy \cite{alipour2019, ahmadi2019a} and is thus motivated by the classical definition of heat \cite{reichl2016}. The von Neumann entropy, $S_t=-\Tr(\rho_t \ln\rho_t)$, varies as $\dot{S}_t = -\sum_{k} \dot{n}_{k,t}\ln{n}_{k,t}$.  That is, the TLS entropy  varies  only with changes in the state eigenvalues, or equivalently, changes in the norm of the Bloch vector.
This  variation also  determines the purity change, which from Eqs.~(\ref{normBlochVector})-(\ref{purity}) readily reads $\dot{\mathcal{P}}_t = \dot{n}_t n_t$ and yields
\begin{equation} \label{EntropyEvolQubit}
\dot{S}_t =\frac{1}{2}\dot{n}_t  \ln\left(\frac{1-n_t}{1+n_t}\right)=
 \frac{\dot{\mathcal{P}}_t}{ 2 n_t} \ln\left(\frac{1-n_t}{1+n_t}\right) . 
\end{equation}
In the `entropy-based' (EB) approach, 
 a variation of the eigenvalues leads to a change of heat (accompanied with entropy change)  whenever the internal energy does not vanish, {\it cf.} Eq.~\ref{ebheat}.
The variation of the internal energy (\ref{internalenergy}) attributed to heat changes is thus defined as 
\begin{eqnarray}  
\dot{Q}_t &=&\frac{\dot{n}_t}{n_t}U_t= \dot{n}_t h_t \cos\alpha_t
\label{ebheat}\\
&=& \dot{q}_t - n_t h_t \frac{d}{dt}\cos\alpha_t  = \dot{\mathbb{Q}}_t^{wc} -  h_t n_t {\hat{h}}_t \cdot \dot{\hat{n}}_t   . \nonumber
\end{eqnarray} 
It can be verified that this definition of heat is consistent with $\dot{Q}_t  = \sum_{k=\pm}\dot{n}_{k,t} \bra{n_{k,t}}H_t\ket{n_{k,t}}$, which is the form proposed in \cite{alipour2019}. 
The remaining terms in the internal energy change are  assigned to work exchange
\begin{eqnarray} 
\dot{W}_t  &=& n_t \frac{d}{dt} \big( h_t \cos \alpha_t \big)  \label{ebwork} \\
&=& \dot{w}_t + n_t h_t \frac{d}{dt}\cos\alpha_t  = \dot{\mathbb{W}}_t^{wc} +  h_t n_t {\hat{h}}_t \cdot \dot{\hat{n}}_t  \nonumber . 
\end{eqnarray}
 The difference between the EB approach and the conventional method, $\dbar \mathbb{Q}_t^{wc}-\dbar Q_t = h_t n_t {\hat{h}}_t \cdot \dot{\hat{n}}_t $, is path dependent and assigned to environment-induced  `dissipative work' \cite{alipour2019}.

 So the  three considered approaches are equivalent when there is no change in the directions of neither the trajectory, nor the driving. Whenever these unit vectors vary,  the assignation of heat and work becomes approach dependent. The HB and EB approaches are equivalent whenever the system is driven along a  trajectory with constant deviation ($\dot{\alpha}_t=0$).   When the angle varies, the contribution in $\dot{\alpha}_t$ is  associated to either heat (`Hamiltonian-based' framework, with energy as preferred basis) or work (entropy-based approach, trajectory basis used as reference). Note that this contribution does not alter entropy nor purity---{\it cf.} Eq.~(\ref{EntropyEvolQubit}).  From Eq. \eqref{eq:cosalpha}, we see it requires a variation in at least one of the overlaps $\langle E_{k,t} \vert n_{k',t} \rangle$, which is related to  a variation of the system coherence in the energy eigenbasis \cite{baumgratz2014}. A  redefinition of the first law of thermodynamics that splits internal energy change into three contributions (heat, work, and coherence) has been recently proposed \cite{bernardo2020}. 
Figure \ref{fig1}(b) presents a schematic illustration of these different distributions of internal energy changes.  Additionally, in the weak-coupling definitions (i), the variation of  $\alpha_t$ is split into work and heat exchanges, originating from   the variation of the unit vectors $\hat{h}_t$ and $\hat{n}_t$, respectively.

 Next, we use the definition of the  instantaneous inverse temperature of the system recently proposed for nonequilibrium settings \cite{alipour2021}, 
\begin{equation}
\label{eq:temperature}
\beta_t = -\frac{{\rm cov}(H_t, \ln \rho_t)}{(\Delta H_t)^2} = \frac{\cos\alpha_t}{2 h_t}\ln\left( \frac{1-n_t}{1+n_t} \right), 
\end{equation}
with ${\rm cov}(AB)\equiv\Tr(AB)/d - \Tr(A) \Tr(B)/d^2$ and $(\Delta H)^2 = \Tr(H^2)/d - \Tr(H)^2/d^2$, $d$ being the dimension, 
to compute the irreversible entropy. 
The latter is approach-dependent and reads, using the EB heat \eqref{ebheat}, 
\begin{subequations}
\begin{align}
\dot{S}_{{\rm i},t}^{\rm EB} &=\dot{S}_t - \beta_t \dot{Q}_t, \label{Sir_def} \\
&=\frac{1}{2}\dot{n}_t  \sin^2 \alpha_t\ln\left(\frac{1-n_t}{1+n_t}\right) \label{Sir_EB}. 
\end{align}
\end{subequations}
In the example below, we compare this result  with the other approaches, that give the irreversible entropy as  $\dot{S}_{{\rm i},t}^{\rm HB}=\dot{S}_t - \beta_t \dot{q}_t$ and $\dot{S}_{{\rm i},t}^{wc}=\dot{S}_t - \beta_t \dot{\mathbb{Q}}_t^{wc}$.

\section{Application to a periodically driven open atom} \label{ModelDynamics}
Let us now compute these definitions in a specific model that represents a microscopic heat pump powered by a laser. This example consists of a two-level atom periodically driven by a classical laser field and interacting with both a photon bath and a dephasing bath. The two baths can have different temperatures and have different interactions with the system: the first is diagonal in the system basis while the second is purely off-diagonal, thus causing  decoherence only with no population transition. Such a model is adequate to describe different physical scenarios \cite{szczygielski2013} including a quantum dot interacting with acoustic phonons \cite{li1999}, an atom driven by an optical field and immersed in a buffer gas \cite{vogl2009}
 or also a driven two-level molecule  with variable dephasing of thermal origin \cite{grandi2016}. 
 This open driven system has been considered and solved in e.g.  \cite{szczygielski2013,gasparinetti2014,bulnescuetara2015, elouard2017a}. We recast below the main points of the derivation, with details in the Appendix, to obtain the quantities relevant for the thermodynamics analysis.
 Note that a similar model with the photon bath only has been looked at using the Bloch equations \cite{elouard2020}, and that the standard thermodynamics approach has been investigated in a periodically driven qubit with purely off-diagonal bath using the stochastic Schr\"odinger equation \cite{donvil2018}. Here, we  combine the two kinds of bath and consider the recently proposed  entropy-based formulation of thermodynamics  \cite{alipour2019, ahmadi2019a}.

The system is an atom  driven by a monochromatic classical field with Hamiltonian 
\begin{equation} \label{Hamiltonian}
H_S(t) = \frac{\omega_0}{2} \sigma_z + \varepsilon (e^{i \Omega t} \sigma_- + e^{- i \Omega t} \sigma_+),
\end{equation}
where $\varepsilon = \varepsilon^*$ is proportional to the laser intensity and $\sigma_{+}=\ket{e}\bra{g}=\sigma_{-}^{\dagger}$ are the atomic transition operators.  
This `renormalized' Hamiltonian includes the Lamb shift, so $H_S(t) = \vec{h}_t \cdot \vec{\sigma}$ with $\vec{h}_t = (\varepsilon \cos(\Omega t),\varepsilon \sin(\Omega t), \frac{\omega_0}{2})$ of constant norm $h_t=\sqrt{\varepsilon^2 + (\omega_0 /2)^2}\equiv h_0$.  Since $H_t^{eg} =\varepsilon e^{-i \Omega t}$, we have  $\Theta_t = \Omega t$ here. We take $\hbar = k_B=1$. 

The internal energy, defined in Eq.  (\ref{internalenergy}), reads, for a general state represented by the density matrix (\ref{rhoGeneral}), 
\begin{equation} \label{eq:intenergymodel}
U_t=  \frac{\omega_0}{2}\Delta_t + 2 \varepsilon |{\rho}_t^{eg}| \cos(\varphi_t - \Omega t).
\end{equation}
Since the norm of the driving Hamiltonian is constant, $\dot{h}_t=0$, regardless of the trajectory,  the  Hamitonian-based definition of work variation (\ref{standardwork}) always vanishes. Indeed,  $\dot{w}_t=0$, so  the internal energy variation is completely assigned to heat change, $\dot{U}_t = \dot{q}_t$. 
 In turn, 
  the conventional  (\ref{AlickiWork}) and  the entropy-based  (\ref{ebwork}) definitions yield non-vanishing work change, that respectively read  $\dot{\mathbb{W}}_t^{wc} = 2 \varepsilon \Omega |\rho_t^{eg}| \sin(\varphi_t - \Omega t) $ and $\dot{W}_t = - \dot{\alpha}_t n_t h_0  \sin \alpha_t$.  The heat variation is therefore reduced by the same quantity such that the variation of internal energy matches in  all approaches.    Difference between the approaches arise  due to variations of the unit vectors defining the direction of the trajectory and driving, $\dot{\hat{n}}_t$ and $\dot{\hat{h}}_t$, respectively.
In the following, we solve the dynamics for a dissipative system  before presenting numerical results for heat and work. 
\\
\\
\textit{Model of the baths and dynamics.}  
Consider the TLS interacts with an environment $H_B = H_z  + H_x$ formed by two baths of harmonic oscillators $H_z = \sum_k \omega_k b\dg_{z,k} b_{z,k}$ and $H_x = \sum_k \omega_k b\dg_{x,k} b_{x,k}$. 
The interaction is divided into a purely dephasing term  $(j=z)$ that is diagonal in the atom basis, and an electromagnetic bath  of photon  $(j=x)$ that is purely off-diagonal. Namely, the interaction  Hamiltonian reads
\begin{equation} \label{eq:V}
V = V_z + V_x = \lambda_{z} \sigma_z \otimes B_z + \lambda_{x} \sigma_x \otimes B_x
\end{equation}
with the bath operators $B_j = \sum_k   g_{j,k}  (b\dg_{j, k} + b_{j, k})$ 
 where  the couplings $g_{j,k}$ relate to the spectral density  and $\lambda_{j}$ 
 accounts for a global coupling strength. The baths participate to the dynamics through the Fourier transform of their  correlation functions,  defined for positive frequency from $G_j(\omega) = \int_{-\infty}^\infty d\tau e^{i \omega \tau} \Tr(B_{j}\dg(\tau) B_{j} \rho_{j})  = e^{-\beta_j \omega}G_j(-\omega)$, and   that fulfills detailed balance since the baths are considered at equilibrium with a thermal density reading $\rho_{j} = e^{- \beta_{j} H_{j}} / \Tr(e^{- \beta_{j} H_{j}})$. 
These correlation functions, together with the bath coupling strengths and the laser parameters, determine the  decay rates $\Gamma_1$, $\Gamma_2$ of two different decay channels, as detailed in the Appendix and below. 

 The system is driven periodically, $H_{S}(t)=H_{S}(t+T)$, with a period $T = 2\pi/\Omega$. Using Floquet theorem \cite{shirley1965,zeldovich1967,szczygielski2021}, the evolution operator  $U_S(t) = \mathcal{T} e^{- i \int_0^t H_S(t')dt'}$, where $\mathcal{T} $ is the time-ordering operator,  can be decomposed  into a periodic operator $P_t$ and a time-independent `average Hamiltonian', denoted $\bar{H}$. It can be verified through time differentiation that  $U_S(t) = P_t e^{- i \bar{H} t}$, 
where $P_t = e^{- it \frac{\Omega}{2}\sigma_z}= P_{t+2T}$ and $\bar{H} = \frac{\delta}{2} \sigma_z + \varepsilon \sigma_x $. Here,  $\delta=\omega_0-\Omega$ is the detuning of the driving laser with respect to the electronic transition and $\Omega_r = \sqrt{4\varepsilon^2 + \delta^2}$ is the Rabi frequency.  Diagonalizing the average Hamiltonian, $\bar{H} = \frac{\Omega_r}{2} \bsigma_0$ where $\bsigma_0  \equiv \bsigma_z= \ket{\bar{e}}\bra{\bar{e}} - \ket{\bar{g}}\bra{\bar{g}}$, yields the `Floquet basis' $( \ket{\bar{e}}, \ket{\bar{g}})\dg = M (\ket{e}, \ket{g})\dg $ with 
\begin{equation} \label{eq:barbasis}
M = \begin{pmatrix}
 \cos\theta &   \sin\theta \\
- \sin\theta &   \cos\theta 
\end{pmatrix}
,
\end{equation} 
where $\cos(2 \theta) = \delta /  \Omega_r$ and $\tan(2 \theta) = 2 \varepsilon/\delta$. The differences between eigenvalues  define the set of `quasi-Bohr frequencies' $\bar{\Omega}_B = \Omega_r \Lambda$ with $\Lambda = \{-1,0,1\}$.
 The system operators $\sigma_j$ involved in the coupling (\ref{eq:V}) evolve, in the driven-system interaction picture $ \sigma_j(t) =  U_S\dg(t) \sigma_j U_S(t)$, according to the quasi-Bohr frequencies  and the driving frequency. This  is clear from the Fourier decomposition 
\begin{equation} \label{fourier}
\sigma_{j}(t)=\sum_{q,p \in \Lambda} e^{i(q\Omega_r+ p \Omega)t} s_{q,p}^{(j)} \bar{\sigma}_q,
\end{equation}
where the real coefficients $s_{q,p}^{(j)}$ are detailed in the Appendix. We denote $\bar{\sigma}_q = M \sigma_q M\dg$ for $q\in\{-,0,+\}$  the Pauli matrices in the Floquet basis. 

\textit{Master equation and resolution.} 
The dynamics is first written in the total interaction picture defined from the evolution with no interaction, $U(t) = U_S(t) e^{- i H_Bt}$, in which the reduced density of the system is denoted  $\tilde{\rho}_t$.
 Assuming weak coupling, the master equation reads 
\begin{eqnarray}\label{eq:MEint2}
\frac{d\tilde{\rho}_t}{dt} &=& {-}\hspace*{-1.5ex}{\sum_{j=\{z,x\}}}\hspace*{-1.5ex}\Tr_{B_j}\hspace*{-1ex}\int \limits_0^{\infty} d\tau [V_j(t), [V_j(t-\tau), \tilde{\rho}_t\otimes \rho_j ]] \nonumber \\  
&=& \mathcal{D}(\tilde{\rho_t}), 
\end{eqnarray}
with  $V_j(t) = U\dg(t) V_j U(t)=\lambda_j \sigma_j (t) \otimes B_j (t)$ and 
 $B_j(t) = e^{ i H_j t} B_j e^{- i H_j t}$. This form allows to get the dissipator and group together all time-dependent terms so as to perform the rotating wave approximation, that leads to a compact dissipator of Lindblad form
\begin{equation} \label{eq:Dlindblad}
\mathcal{D}(\tilde{\rho}_t) = \sum_{q \in \Lambda} \gamma_q \left( \bar{\sigma}_q \tilde{\rho}_t \bar{\sigma}\dg_q - \frac{1}{2} \{\bar{\sigma}\dg_q \bar{\sigma}_q, \tilde{\rho}_t \}\right). 
\end{equation}
The Lindblad operators thus correspond to the Pauli matrices in the Floquet basis. 
The relaxation rates account for the two baths through $ \gamma_q =  \gamma_q^{(x)} +  \gamma_q^{(z)}$, defined from  $ \gamma_q^{(j)} = \lambda^2_j \sum_{p\in \Lambda}(s^{(j)}_{q,p})^2G_j\big(-q \Omega_r - p\Omega \big) $. 

The master equation \eqref{eq:MEint2} is first solved in the interaction picture, with the density matrix obtained in the Floquet basis. We then recast the density matrix in the Schr\"odinger picture and express it in the atom basis---details are given in the Appendix. The evolution of the quantum state
\begin{equation}\label{eq:rhot}
\rho_t = \frac{1}{2}(\mathbb{1} + \vec{N}_t \cdot P_t \vec{\bar{\sigma}} P\dg_t)
\end{equation}
is set by the elements of the Bloch vector $\vec{N}_t = (X_t,Y_t,Z_t)$ with 
\begin{subequations}\label{sol_rhot}
\begin{align}
X_t&=2 \, e^{- \Gamma_2 t} {\rm Re} \big( e^{- i \Omega_r t }\rho_0^{\overline{eg}}\big),\\
Y_t&=  - 2\, e^{- \Gamma_2 t} {\rm Im} \big( e^{- i \Omega_r t }\rho_0^{\overline{eg}}\big), \\
Z_t&=  \bar{\Delta}_t= e^{-\Gamma_1 t} \Big(\bar{\Delta}_0 + 2\kappa \Big)   - 2\kappa . 
\end{align}
\end{subequations}
The initial population inversion in the Floquet basis reads $\bar{\Delta}_0 = \Delta_0 \cos(2\theta)+2\text{Re}(\rho_0^{eg})\sin(2\theta)$ and the coherence term reads $\rho_0^{\overline{eg}} = -\frac{\Delta_0}{2} \sin(2\theta)+\text{Re}(\rho_0^{eg})\cos(2\theta)+i\text{Im}(\rho_0^{eg})$. 
The population inversion and the norm of the Bloch vector evolve as   
\begin{subequations}
\begin{align}
\Delta_t &=Z_t \cos2 \theta - X_t \sin 2\theta  \label{DeltaModel},  \\
n_t^2 &= X_t^2 + Y_t^2 + Z_t^2 = e^{- 2 \Gamma_2 t} 4 |\rho_0^{\overline{eg}}|^2 + \bar{\Delta}_t^2. 
\end{align}
\end{subequations}
 The effects of the baths appear in the decay rates
\begin{subequations}
\begin{align}
\Gamma_1 &= \gamma_+ + \gamma_- = \gamma_+^{(z)} + \gamma_-^{(z)}+\gamma_+^{(x)} + \gamma_-^{(x)}, \\
\Gamma_2 &= \frac{\Gamma_1}{2} + 2 \gamma_0, 
\end{align}
\end{subequations} and through the dimensionless constant $\kappa = \frac{1}{2}\frac{\gamma_- - \gamma_+}{\gamma_- + \gamma_+}$, that is  related to steady-state values and bounded as $|\kappa| \leq \frac{1}{2}$. 
 Large values of $\gamma_{\pm}$ lead to large decay rates for the two channels,  leading to fast exponential decay of the Bloch vector coordinates. In turn, $\gamma_0$ only modulates the second decay channel and does not necessarily yields a fast decay.

The state coherence follows from Eqs.~(\ref{eq:rhot})-(\ref{sol_rhot}) as 
\begin{equation}
\rho_t^{eg} = \frac{1}{2} \big(X_t \cos 2\theta - i Y_t + Z_t \sin 2\theta \big)e^{-i \Omega t}. \label{fig:rhoeg}
\end{equation}
It also decays exponentially with time following the decay of the Bloch vector coordinates. 
From $\rho_t^{eg} \equiv |\rho_t^{eg}| e^{- i\varphi_t}$, the coherent term gives the relative angle between the vectors characterizing the driving Hamiltonian and the state in the ($xy$)-plane, $\varphi_t - \Theta_t$, satisfying    
\begin{equation} \label{eq:xyangle}
\cos(\varphi_t-\Omega t) = \frac{1}{2  |\rho_t^{eg}| }\left( Z_t \sin 2\theta + X_t \cos 2\theta \right). 
\end{equation}
Substituting Eqs. (\ref{DeltaModel}) and (\ref{eq:xyangle}) into Eq. (\ref{eq:intenergymodel}) yields
\begin{equation}
U_t = Z_t \big(\frac{\omega_0}{2}\cos 2\theta + \varepsilon \sin 2\theta \big) + X_t \big(\varepsilon \cos 2\theta -\frac{\omega_0}{2}  \sin 2\theta \big). \label{UtModel}
\end{equation}
The  decay of the  population inversion, coherence, and angle is a bi-exponential
 with rates dictated by the laser intensity, the bath coupling strengths and correlation functions---see Eq. (\ref{eq:rates}) for the  explicit expressions. 
In the steady state (SS), Eq. (\ref{sol_rhot}) gives $X_{\rm ss}=Y_{\rm ss}=0$ and $Z_{\rm ss}=-2\kappa$. The population inversion becomes $\Delta_{\rm ss}=-2\kappa\cos2\theta$. The state coherence oscillate at the driving frequency, namely $\rho_{\rm ss}^{eg}=-\kappa\sin2\theta e^{-i\Omega t}$, and $\vec{h}_{\rm ss}^{xy}$ and $\vec{n}_{\rm ss}^{xy}$, that denote the vectors in the ($xy$)-plane,  rotate in phase. Consequently, the cosine on the l.h.s of (\ref{eq:xyangle}) is constant, $\cos(\varphi_{\rm ss}-\Omega t)=-\text{sign}(\kappa\sin2\theta)$. In addition, as the $z$-components of both $\vec{n}$ and $\vec{h}$ are then constant,  the angle between the two vectors $\alpha_{\rm ss}$ is also constant.  The norm of the Bloch vector reaches the SS value of $n_{\rm ss}=2|\kappa|$, which is independent of $\gamma_0$ and grows with the absolute value of the difference between $\gamma_+$ and $\gamma_-$. This feature is translated in the SS values of purity, $\mathcal{P}_{\rm ss} = 2\kappa^2 +1/2$, and entropy, $S_{\rm ss}=\ln\frac{2}{\sqrt{1-4\kappa^2}}+ |\kappa|\ln\frac{1-2|\kappa|}{1+2|\kappa|}$. A similar behavior is found for the steady-state internal energy, that reads $U_{\rm ss}=-\kappa (2\varepsilon\sin 2\theta+\omega_0 \cos 2\theta)$, but accounts for the sign of the difference between $\gamma_+$ and $\gamma_-$.

\begin{figure}
	\includegraphics[width=\columnwidth]{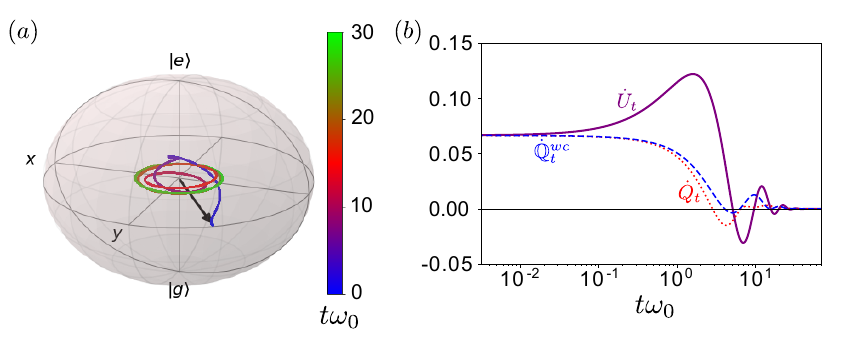}
\caption{{\bf Time evolution} of (a) the Bloch vector $\vec{n}_t$ representing the reduced density matrix and (b) its corresponding thermodynamic quantities. The system is initialized in a thermal state at inverse temperature $\beta \omega_0 =1$ and evolves according to Eqs.~(\ref{DeltaModel})-(\ref{fig:rhoeg}) until reaching the steady state, where its representative  vector oscillates in the ($xy$)-plane. The color bar shows the time evolution. 
	 The variation of internal energy (purple) directly gives  the semiclassical, HB heat in this example. During the transient, it differs from the conventional definition (dashed blue) and the entropy-based approach (dotted red). 
	The laser is tuned resonantly  with the atom transition, $\Omega = 1$, with intensity $\varepsilon=0.3$.  Decay rates are fixed to $\gamma_+ = 0.1$, $\gamma_{-}=0.05$ and $\gamma_0 = 0.05$. All quantities are in units of $\omega_0$.
\label{fig:thermo_t}}
\end{figure}

{\it Numerical simulations and discussion.} 
 From the resolution of the dynamics \eqref{sol_rhot}, 
it is  straightforward to compute the variation of  thermodynamics properties of the driven open TLS  using the time derivatives
\begin{subequations}
\begin{align}
\dot{X}_t &= -\Gamma_2 X_t -\Omega_r Y_t, \\
\dot{Y}_t &= -\Gamma_2 Y_t + \Omega_r X_t, \\
\dot{Z}_t &= -\Gamma_1 (Z_t + 2\kappa).  
\end{align}
\end{subequations}
 The conventional approach (i) defined in Eqs.  (\ref{Alicki}) gives the variation of work as $\dot{\mathbb{W}}_t^{wc} = \varepsilon \Omega Y_t $. Then, the Hamiltonian-based approach (ii), Eq. \eqref{standard},  gives, as mentioned,  zero work. So all changes of internal energy are directly assigned to heat and are straightforward  from \eqref{UtModel}. Finally, the entropy-based approach (iii), gives heat and work from Eqs.~\eqref{ebheat}-\eqref{ebwork}.

\begin{figure}
	\includegraphics[width=\columnwidth]{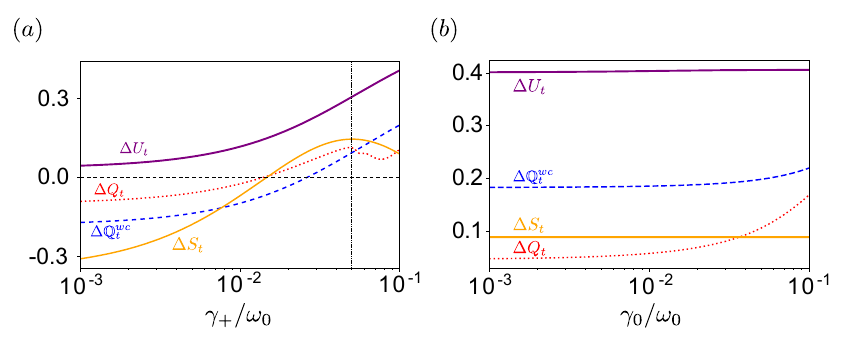}
	\caption{
	{\bf Variation of thermodynamic quantities} as function of the  decay rates (a) $\gamma_+  = \gamma_+^{(x)} +  \gamma_+^{(z)}$ with $\gamma_0=0.05$ and (b) $\gamma_0=\gamma_0^{(x)}$ with $\gamma_+=0.1$, for fixed $\gamma_-=0.05 $ in both cases. The variation of internal energy (purple)  directly represents variation of semiclassical heat here.  It differs from the variation of heat according to the `conventional' (dashed blue) and `entropy-based' (dotted red) approaches, see text. The variation of entropy is also plotted (orange). For each quantity, we plot the net variation, i.e., the change  integrated from initial state to onset of steady state ($t_{\rm ss} =30/\omega_0$). 	The atom and laser parameters are as in Fig.~\ref{fig:thermo_t}  and all quantities are in units of $\omega_0$. \label{fig:thermo}}
\end{figure}

In order to compare  the  three considered thermodynamics approaches, we present the evolution  from an initial thermal state in Fig. \ref{fig:thermo_t} and show the dependence of the thermodynamic quantities on the   decay rates in Fig. \ref{fig:thermo}. Numerical applications for other initial states are presented in the Appendix.  
Fig.~\ref{fig:thermo_t}a illustrates  the time evolution of the Bloch vector $\vec{n}_t$, computed from the coherence (\ref{fig:rhoeg}) and the population inversion (\ref{DeltaModel}), 
 that decays exponentially in time up to the steady state where it oscillates in time in the ($xy$)-plane, as predicted analytically. 
Fig.~\ref{fig:thermo_t}b shows the evolution of  the exchange of heat according to three approaches. All these variations vanish at steady state. The evolution of internal energy changes (purple), which also corresponds to the semiclassical, `Hamiltonian-based' heat exchange, drastically differs from the other two approaches: it first increases until reaching a maximum before decreasing and oscillating toward the steady state, zero value. In turn, the `conventional' (dashed blue) and `entropy-based' (dotted red) exchanges of heat evolve similarly, decreasing from the initial time and then oscillating before reaching the steady state. The HB approach does not predict production of work. As commented,  the difference between the conventional and EB heat is understood as a `dissipative' contribution to work \cite{alipour2019}.

Figure~\ref{fig:thermo}a shows the entropy changes (orange) integrated over the full transition (from initial time to onset of steady state, i.e.  $\Delta S = \int_0^{t_{\rm ss}} \dot{S}_\tau d\tau$) as function of the relaxation rate  $\gamma_{+}$. It has an extremum at  $\gamma_+ = \gamma_-$  (vertical line in Figure~\ref{fig:thermo}a),  around which point it is  symmetric. At this point, the steady state is maximally mixed and  the EB heat  shows a local peak. That is consistent with the fact that $S_{\rm ss}$ depend on $|\kappa|=|\gamma_+ - \gamma_-|$---the same behavior applies to the purity, not shown here. Additionally, the variation of internal energy grows with $\gamma_{+}$, as expected since $U_{\rm ss}\propto (\gamma_{+}-\gamma_{-})$, so the variation of HB heat also does.  The conventional heat follows a similar trend with a quasi-constant offset. However, the variation of EB heat grows up to $\gamma_+ \approx \gamma_-$, after which point it  stops being monotonic, which contrasts with the behavior of the other  approaches. 
Fig.~\ref{fig:thermo}b shows the dependency of the thermodynamics quantities as function the  decay rate $\gamma_0$. As expected, there is no variation of entropy and internal energy.  So HB heat is constant. However, the EB approach predicts a larger increase in the variation of heat as function of this relaxation rate than the other models. 
This heat increases at the expense of decreasing the `dissipative work', which originates from the coherence part of the dynamics \cite{alipour2019}. As mentioned above, it is the energy changes due to the coherent part of the dynamics that make the difference and are either assigned to work (EB) or heat (conventional approach). 
This figure also shows that the  considered three approaches lead to very different characterizations of the thermodynamics evolution of this system over a wide range of  decay rates.

\begin{figure}
	\includegraphics[width=\columnwidth]{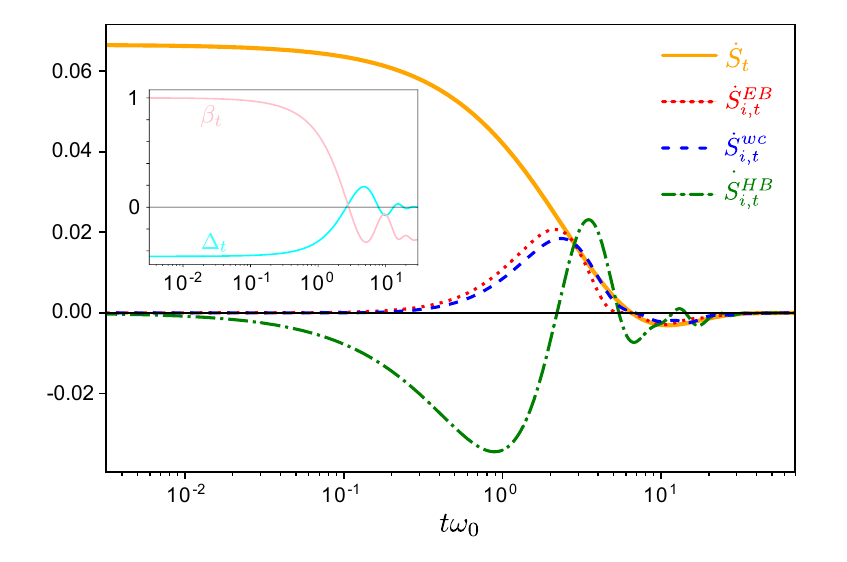}
	\caption{ Time-evolution of the irreversible entropy as obtained from the different approaches, Eq. \eqref{Sir_def} and below, for the initial thermal state with parameters  as in Fig. \ref{fig:thermo_t}. The inset shows the evolution of the system instantaneous inverse temperature, Eq. \eqref{eq:temperature}, and population inversion $\Delta_t$.  All quantities are in units of $\omega_0$.\label{fig:Sirr}}
\end{figure}

 Finally, Fig.~\ref{fig:Sirr} presents the evolution of the entropy and irreversible entropy according to the different approaches. The inset illustrates the evolution of the instantaneous inverse temperature (pink), according to Eq. (\ref{eq:temperature}), and the population inversion (cyan). At short times, the ground state is more populated than the excited state (i.e. $\Delta_t > 0$) and the temperature is positive---as expected from the populations. The conventional and EB irreversible entropy (dashed blue and dotted red, respectively) are non-negative and exhibit very similar trends because of similar heat predictions---{\it cf} Fig.~\ref{fig:thermo_t}(b). In turn, the HB irreversible entropy (dot-dashed green) starts decreasing and being negative, until reaching a minimum. At the crossing $\beta_t=0$, the irreversible entropy coincides with the von Neuman entropy, regardless of the approach. At larger times, the temperature becomes negative while the population inversion oscillates toward its steady state, zero value, as expected for resonant excitation of the atom---note that states with negative temperature are physically relevant \cite{puglisi2017} and have been experimentally demonstrated \cite{medley2011,braun2013}. Simultaneously, irreversible entropy decays to zero for the three approaches. However, the decay of HB irreversible entropy is accompanied by sharp oscillations with positive and negative values, while conventional and EB approaches predict a softer decay, without oscillations, and  reaching negative values only at final times. We remark that the conventional and EB predictions are very similar, as expected from the heat rates, with the small difference asigned to production of `dissipative' work.

 Note that the  three approaches coincide only for very specific conditions, such as when the system is initialized in a maximally mixed state---see Appendix for details.

\section{Conclusion}

 We presented the thermodynamics of a two-level system on the Bloch sphere, focusing on three selected approaches to assign the change of internal energy into  quantum heat or work:  one  approach motivated by work as changes in the system energy,  another motivated by heat as changes in the system entropy,  and the one conventionally used today. The two contributions that can switch from  heat to  work between different approaches originates from variations in the direction of the trajectory or the driving Hamiltonian. 
Changes in the relative directions---directly given by the variation of the angle  between the respective vectors on the Bloch sphere---can be interpreted as purely quantum and are physically related to the system quantum coherence in the energy basis. 

We solved the dynamics of a microscopic heat pump powered by laser to illustrate the differences. Even in the case of weak-coupling and Markovian evolution, the considered approaches lead to different assignation of heat and work. This is because an open quantum system has not a unique Lindblad description and work can still be extracted from the open, dissipative part. 
 With this in mind, we still find that the semiclassical, HB approach predicts results very distant from the conventional approach, developed in the weak-coupling regime that we are considering. The EB approach in turn shows  small differences which can be interpreted as a  corrections  to the conventional approach emerging from the dissipative work contribution. We also find larger discrepancies on  the HB irreversible entropy, with   sharper oscillations and negative values over a longer time.
Considering that the differences between approaches mainly come from the system coherence in the energy basis, experiments with pure dephasing could provide more intuition in terms a definite recommendation.

\emph{Acknowledgments.---}We thank  S. Alipour, A. Rezakhani, A. del Campo and J. Yang for  discussions and  comments on the manuscript. 


%


\appendix 

\onecolumngrid

\section{Details for the derivation of the model dynamics and its resolution \label{MEapp}} 
We recast below the main points for the derivation of the master equation and its resolution.

{\it Floquet decomposition}. The system is driven periodically, $H_{S}(t)=H_{S}(t+T)$ with a period $T = 2\pi/\Omega$. 
The evolution with periodically-driving Hamiltonians can be obtained using Floquet theorem \cite{shirley1965,zeldovich1967,szczygielski2021}. In that case, the evolution operator $U_S(t) = \mathcal{T} e^{- i \int_0^t H_S(t')dt'}$, where $\mathcal{T} $ is the time-ordering operator,  can be decomposed  into a periodic operator $P_t$ and a time-independent `average Hamiltonian', denoted $\bar{H}$. It can be verified through time differentiation that  
\begin{equation} \label{us}
U_S(t) = P_t e^{- i \bar{H} t}
\end{equation}
where $P_t = e^{- it \frac{\Omega}{2}\sigma_z}= P_{t+2T}$ and $\bar{H} = \frac{\delta}{2} \sigma_z + \varepsilon \sigma_x $.  The average Hamiltonian is diagonalized as $\bar{H} = \frac{\Omega_r}{2} \bsigma_0$ in the eigenbasis
\begin{equation} \label{eq:barbasis}
\begin{split}
\ket{\bar{e}}&= \cos\theta \ket{e} +  \sin\theta \ket{g}, \\
\ket{\bar{g}}&= - \sin\theta \ket{e} +   \cos\theta \ket{g}, 
\end{split}
\end{equation} 
with $\cos(2 \theta) = \frac{\delta}{ \Omega_r}$ and $\tan(2 \theta) = \frac{2 \varepsilon}{\delta}$. The eigenstates define the `Floquet basis', while the differences between eigenvalues  define the set of `quasi-Bohr frequencies' $\bar{\Omega}_B = \Omega_r \Lambda$ with $\Lambda = \{-1,0,1\}$. 

Let us first look at the evolution, in the driven-system interaction picture, of the system operators $\sigma_j$ involved in the coupling, i.e., $ \sigma_j(t) =  U_S\dg(t) \sigma_j U_S(t)$. The first one easily follows from Eqs. (\ref{us}) and (\ref{eq:barbasis}) as 
\begin{eqnarray}
\sigma_z(t)  &=& e^{ i t \frac{\Omega_r}{2}  \bar{\sigma}_0 } \sigma_z  e^{- it \frac{\Omega_r}{2} \bar{\sigma}_0  } \\
&=& \cos(2\theta) \bar{\sigma}_0 - \sin(2\theta) (e^{ i  \Omega_r t} \bar{\sigma}_+  + e^{ -i  \Omega_r t} \bar{\sigma}_-) , \nonumber
\end{eqnarray}
where the evaluation in the last line follows from the BCH formula and the commutator 
$[\bar{\sigma}_0, \sigma_z] =  \frac{4 \varepsilon}{\Omega_r} (\bar{\sigma}_- - \bar{\sigma}_+) $. The second operator of interest is 
\begin{equation}
\begin{split}
\sigma_x(t)  &=  e^{ i \frac{\Omega_r}{2}  \bar{\sigma}_0 t} (e^{i\Omega t}\sigma_{+} + e^{-i\Omega t}\sigma_-) e^{- i \frac{\Omega_r}{2}  \bar{\sigma}_0 t}  \\
&= e^{it \Omega} \Big(
\frac{\sin(2\theta)}{2} \bar{\sigma}_z 
+\frac{\cos (2\theta){+}1}{2} e^{ i \Omega_r t} \bar{\sigma}_+ {+}\frac{\cos (2\theta){-}1}{2} e^{- i \Omega_r t} \bar{\sigma}_-\Big) {+} \rm{h.c.},\\
\end{split}
\end{equation}
as readily  follows from evaluating $P_t^{\dagger}\sigma_z P_t$ thanks to the BCH formula and the commutator
$[\bar{\sigma}_z, \sigma_+] = 2 (\cos^2 \theta \bar{\sigma}_+ + \sin^2 \theta \bar{\sigma}_-)$. The time evolution is thus dictated by the quasi-Bohr frequencies  and the driving frequency. Indeed, it can be recast into the Fourier decomposition 
\begin{equation} \label{fourier}
\sigma_{j}(t)=\sum_{q,p \in \Lambda} e^{i(q\Omega_r+ p \Omega)t} s_{q,p}^{(j)} \bar{\sigma}_q , 
\end{equation}
with the real coefficients  $s^{(z)}_{0,0} = \cos(2\theta) $ and $s^{(z)}_{\pm,0}=-\sin(2\theta)$ and $s_{q,\pm}^{(z)}=0$  for the dephasing bath and   $s_{0,\pm}^{(x)}=\frac{1}{2} \sin(2\theta)$, $s_{\pm,\pm}^{(x)}=\frac{1}{2} (\cos(2\theta)+1)$, $s_{\pm,\mp}^{(x)}=\frac{1}{2} (\cos(2\theta)-1)$  and $s_{q,0}^{(x)}=0$ for the photon bath. 

{\it Master Equation.} Let us now look at the master equation of the reduced system in the total interaction picture defined from the evolution with no interaction $U(t) = U_S(t) e^{- i H_Bt}$, and in which the reduced density matrix is denoted with a tilde, $\tilde{\rho}_t$. 
The von-Neumann equation for the total density matrix  $\varrho_t$ reads $\frac{d \tilde{\varrho}_t}{dt} = - i [V(t),  \tilde{\varrho}_t]$ and leads, assuming the Born-Markov approximation $ \tilde{\varrho}_t = \tilde{\rho}_t \otimes \rho_B$ and  uncorrelated baths, to the master equation for the reduced  density matrix
\begin{eqnarray}\label{eq:MEint}
\frac{d\tilde{\rho}_t}{dt} = {-} {\sum_{j=\{z,x\}}}\Tr_B\int_0^{\infty} d\tau [V_j(t), [V_j(t-\tau), \tilde{\varrho}_t]]  
= \mathcal{D}_z (\tilde{\rho}_t) + \mathcal{D}_x (\tilde{\rho}_t). 
\end{eqnarray}
The interacting Hamiltonian, in the interaction picture, reads  $V_j(t) = U\dg(t) V_j U(t)=\lambda_j \sigma_j (t) \otimes B_j (t)$,
with $B_j(t) = e^{ i H_j t} B_j e^{- i H_j t}$. This form allows to get the dissipator and group together all time-dependent terms so as to perform the rotating wave approximation, that leads to the compact Lindblad form
\begin{equation} \label{eq:Dlindblad}
\mathcal{D}_{j}(\tilde{\rho}_t) = \sum_{q \in \Lambda} \gamma_q^{(j)} \left( \bar{\sigma}_q \tilde{\rho}_t \bar{\sigma}\dg_q - \frac{1}{2} \{\bar{\sigma}\dg_q \bar{\sigma}_q, \tilde{\rho}_t \}\right), 
\end{equation}
provided that  $G^*(-\omega) = G(\omega)$. 
The Lindblad operators thus correspond to the Pauli matrices in the Floquet basis. 
The relaxation rates are defined by $ \gamma_q^{(j)} = \lambda^2_j \sum_{p\in \Lambda}(s^{(j)}_{q,p})^2G_j\big(-q \Omega_r - p\Omega \big) $. 
Specifically, the  rates for the diagonal coupling involve dephasing only $\gamma_\pm^{(z)} =  \lambda^2_z \sin^{2}(2\theta) G_z(\mp \Omega_r)$, since $\gamma_0^{(z)} =\lambda^2_z \cos^{2}(2\theta) G_z(0) $ is zero. 
In turn, the rates from the electromagnetic bath read
$\gamma_{0}^{(x)} = \lambda_{x}^2 \frac{\sin^{2}(2\theta)}{4} (G_{x}(\Omega) + G_{x}(-\Omega))$ and 
\begin{equation}
\begin{split}
\gamma_{\pm}^{(x)} = \lambda_{x}^2 \Big[    \Big( \frac{\cos(2\theta)+ 1}{2} \Big)^2 G_{x}(\mp \Omega_{+})
+ \Big( \frac{\cos(2\theta)- 1}{2} \Big)^2  G_{x}(\pm \Omega_{-}) \Big], 
	\end{split}
	\end{equation}
where $\Omega_{\pm}\equiv  \Omega \pm \Omega_r $. 

{\it Solution of the dynamics.} 
We solve the master equation for the density matrix in the interaction picture. This is equivalent to solving the system in the Floquet basis 
\begin{equation}
\frac{d\tilde{\rho}_t}{dt}= \left[ \begin{array}{cc}
- \gamma_-   \bra{\bar{e}} \tilde{\rho}_t \ket{\bar{e}} + \gamma_+   \bra{\bar{g}} \tilde{\rho}_t \ket{\bar{g}} & -\left( \frac{\gamma_+ + \gamma_-}{2} + 2 \gamma_0\right) \bra{\bar{e}} \tilde{\rho}_t \ket{\bar{g}}	\\
-\left( \frac{\gamma_+ + \gamma_-}{2} + 2 \gamma_0\right) \bra{\bar{g}} \tilde{\rho}_t \ket{\bar{e}}		&	\gamma_-   \bra{\bar{e}} \tilde{\rho}_t \ket{\bar{e}} -  \gamma_+   \bra{\bar{g}} \tilde{\rho}_t \ket{\bar{g}}		
\end{array}
\right].
\end{equation}
Equivalently,
\begin{subequations}
\begin{align}
\frac{d}{dt} \bra{\bar{e}} \tilde{\rho}_t \ket{\bar{e}} &= \gamma_+ - (\gamma_+ +  \gamma_-)   \bra{\bar{e}} \tilde{\rho}_t \ket{\bar{e}},  \\
\frac{d}{dt}  \bra{\bar{e}} \tilde{\rho}_t \ket{\bar{g}} & = - \left( \frac{\gamma_+ + \gamma_-}{2} + 2 \gamma_0\right)   \bra{\bar{e}} \tilde{\rho}_t \ket{\bar{g}}. 
\end{align}
\end{subequations}
Thus, introducing the decay rates $\Gamma_1 = \gamma_+ + \gamma_-$ and $\Gamma_2 = \frac{\Gamma_1}{2}+ 2 \gamma_0$,  we find 
\begin{subequations} \label{tilderho_t}
\begin{align}
\bra{\bar{e}} \tilde{\rho}_t \ket{\bar{e}}  &= \frac{\gamma_+}{\Gamma_1} + \Big( \rho_0^{\bar{ee}} - \frac{\gamma_+}{\Gamma_1} \Big) e^ {-\Gamma_1 t} =\frac{1}{2}   - \kappa+  e^{- \Gamma_1 t} \Big(\frac{\bar{\Delta}_0}{2}+\kappa\Big) \equiv \frac{1}{2} +  \frac{\bar{\Delta}_t}{2}, \\
\bra{\bar{e}} \tilde{\rho}_t \ket{\bar{g}} &= e^{ - \Gamma_2 t }   \bra{\bar{e}} {\rho}_0 \ket{\bar{g}}, 
\end{align}
\end{subequations}
where $\bar{\Delta}_t = \bra{\bar{e}} \tilde{\rho}_t \ket{\bar{e}}-\bra{\bar{g}} \tilde{\rho}_t \ket{\bar{g}}$ and $\kappa=(\gamma_- - \gamma_+)/2(\gamma_- + \gamma_+)$. The later is bounded,  $|\kappa|\leq 1/2$, for positive  correlation functions $G(\omega)$. 

The solution of the density matrix can readily be recast in the Schr\"odinger picture. Its elements in the atom basis read
\begin{eqnarray} \label{rhoee}
\frac{\Delta_t}{2} &=&  \frac{\bar{\Delta}_t}{2}\cos(2\theta)-\text{Re}(\bra{\bar{e}} \tilde{\rho}_t \ket{\bar{g}}e^{-i\Omega_r t}) \sin(2\theta),  \\ 
e^{i \Omega t} \rho_t^{eg}&=& \frac{\bar{\Delta}_t}{2}\sin(2\theta) + \bra{\bar{e}} \tilde{\rho}_t \ket{\bar{g}}e^{-i\Omega_r t} \cos^2\theta - \bra{\bar{g}} \tilde{\rho}_t \ket{\bar{e}}e^{i\Omega_r t} \sin^2\theta \nonumber \\
&=&\frac{\bar{\Delta}_t}{2}\sin(2\theta)  
+\text{Re}(\bra{\bar{e}} \tilde{\rho}_t \ket{\bar{g}}e^{-i\Omega_r t}) \cos(2\theta)+i\text{Im}(\bra{\bar{e}} \tilde{\rho}_t \ket{\bar{g}}e^{-i\Omega_r t}).  \label{rhoeg}
\end{eqnarray} 
The density matrix can be written in a compact form using the Bloch vector, that we give in the main text, Eqs.~(\ref{eq:rhot})- (\ref{sol_rhot}).

For completeness, we give the decay rates using the atom and laser parameters 
\begin{subequations} \label{eq:rates}
\begin{align}
\Gamma_1 & =\frac{1}{4 \Omega_r^2}\{ \lambda_z^2 (4\varepsilon^2)g_{z,+}(\Omega_r) +
\lambda_x^2 (\omega_0 - \Omega_{-})^2 g_{x,+}(\Omega_+)  
+ \lambda_x^2 (\omega_0 - \Omega_{+})^2  g_{x,+}(\Omega_-) \},  \label{Gamma1} \\
\Gamma_2 &=\frac{1}{8 \Omega_r^2}\{\lambda_x^2 (\omega_0 - \Omega_{-})^2  g_{x,+}(\Omega_+) 
+ \lambda_x^2 (\omega_0 - \Omega_{+})^2  g_{x,+}(\Omega_-)  
+(4\varepsilon)^2 (\lambda_x^2 g_{x,+}(\Omega)  + \lambda_z^2 g_{z,+}(\Omega_r))   \}, \label{Gamma2} \\
\kappa &= \frac{1}{8\Omega_r^2 \Gamma_1}\{\lambda_x^{2} (\omega_0 - \Omega_-)^2 g_{x,-}(\Omega_+) - \lambda_x^{2} (\omega_0 - \Omega_+)^2 g_{x,-}(\Omega_-) + \lambda_z^2 (4\varepsilon)^2 g_{z,-}(\Omega_r)\}, \label{kappa}
\end{align}
\end{subequations}
where $g_{j,\pm} (\omega)\equiv G_j (\omega) \pm G_j (-\omega)$. 

\medskip 

\twocolumngrid
\section{Numerical simulations for different initial states}
\begin{figure}
	\includegraphics[width= \columnwidth]{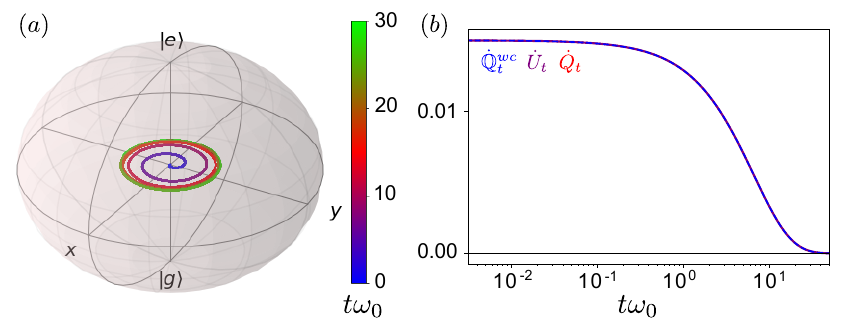}
	\caption{{\bf Initial maximally mixed state.}  Time evolution of the (a) Bloch vector and (b) exchange of heat for the three approaches. The  decay rates are fixed to $\gamma_+=0.1 \omega_0$ and $\gamma_- = \gamma_0 = 0.05\omega_0$. \label{fig:maxmixed_time}} \smallskip \smallskip \smallskip
\end{figure}

\begin{figure}
	\includegraphics[width= \columnwidth]{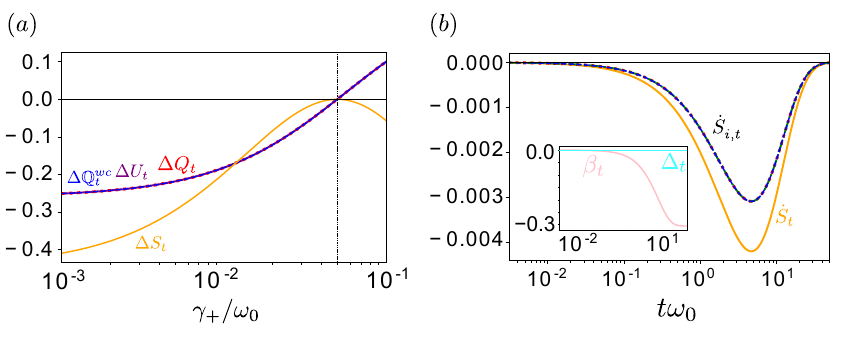}
	\caption{{\bf Initial maximally mixed state.}  (a) Total variation of entropy (yellow) and heat according to the three thermodynamics approaches, from the initial time to the onset of the steady state $t_{ss} \omega_0 = 30$, as function of $\gamma_+$ -- the other  decay rates are fixed again as $\gamma_- = \gamma_0 = 0.05\omega_0$. (b) Evolution of entropy (yellow) and irreversible entropy. The inset shows the evolution of the system instantaneous temperature (pink) and population inversion (cyan). \label{fig:maxmixed_integrated}}
\end{figure}

\begin{figure}
	\includegraphics[width= \columnwidth]{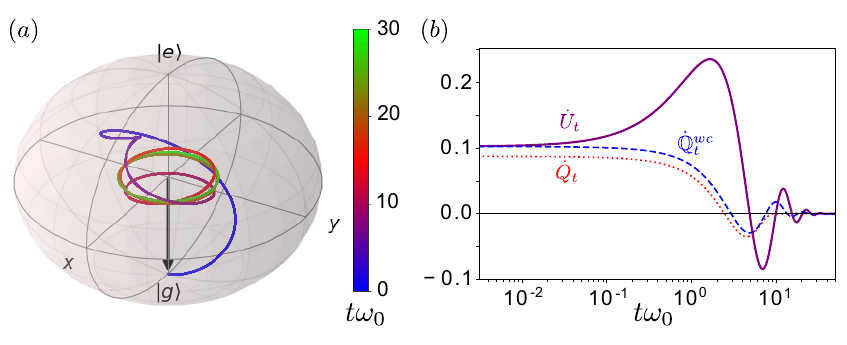}
	\caption{{\bf Initial ground state.}  Time evolution of the (a) Bloch vector and (b) exchange of heat for the three approaches. The  decay rates are fixed to $\gamma_+=0.1 \omega_0$ and $\gamma_- = \gamma_0 = 0.05\omega_0$. \label{fig:ground_time}}
\end{figure}

\begin{figure}
	\includegraphics[width= \columnwidth]{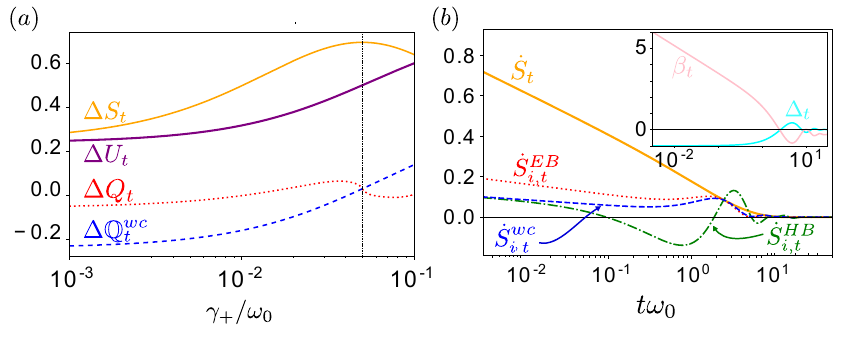}
	\caption{{\bf Initial ground state.}   (a) Total variation of entropy (yellow) and heat according to the three thermodynamics approaches, from the initial time to the onset of the steady state $t_{ss} \omega_0 = 30$, as function of $\gamma_+$ -- the other  decay rates are fixed again as $\gamma_- = \gamma_0 = 0.05\omega_0$. (b) Evolution of entropy (yellow) and irreversible entropy. The inset shows the evolution of the system instantaneous temperature (pink) and population inversion (cyan).   \label{fig:ground_gamma}}
\end{figure}

We show below the results when the system is initialized in the maximally mixed state (Figs.~\ref{fig:maxmixed_time}-\ref{fig:maxmixed_integrated}) or in the pure ground state (Figs.~\ref{fig:ground_gamma}-\ref{fig:ground_time}). 
The dynamics is solved as in the main text, with the same laser parameters, i.e.,  resonant with the atom transition frequency  $\omega_0 =\Omega$ and  with intensity  $\varepsilon = 0.3 \omega_0$.

When the initial state is maximally mixed, the initial population inversion $\Delta_0$ and coherence $\rho_0^{eg}$ are both zero. Eq. \eqref{sol_rhot} leads to $X_t = Y_t = 0$ and $Z_t = 2 \kappa(e^{-\Gamma_1 t}-1)$. The state vector $\vec{n}_t = 2\frac{Z_t}{\Omega_r}(\varepsilon \cos(\Omega t), \varepsilon \sin(\Omega t), \delta/2)$ is thus aligned with the system Hamiltonian $\vec{h}_t = (\varepsilon \cos(\Omega t), \varepsilon \sin(\Omega t), \omega_0/2)$. Since the angle $\alpha_t$ is constant in time, the  HB and EB thermodynamic approaches coincide for this particular initial state and, thus, the variation of internal energy is fully assigned to heat exchange  in both approaches. Besides, the `conventional' production of work is given, according to Eq. (\ref{AlickiWork}), as $\dot{\vec{h}}_t \cdot \vec{n}_t$. Since $\dot{\vec{h}}_t = \epsilon \Omega (-\sin(\Omega t), \cos(\Omega t), 0)$, hence the `conventional work' also vanishes and, thus, the three thermodynamics approaches agree for the maximally mixed initial state, as illustrated in Figs.~\ref{fig:maxmixed_time} and ~\ref{fig:maxmixed_integrated}. 

In Fig.~\ref{fig:maxmixed_time}(a) we observe the exponential decay to the SS in the $z=0$ plane during all the transient, i.e. the state remains maximally mixed along the evolution. In Fig.~\ref{fig:maxmixed_time}(b) the heat variation---identical for the three approaches---shows a monotonic decay, without oscillations when reaching the SS or a maximum for the variation of internal energy, as found for the thermal state in Fig.~\ref{fig:thermo_t}(b).

Additionally, since $U_0 = 0$, hence $\Delta U = \int_{0}^{t_{\rm ss}}\dot{U}_{\tau}d\tau = U_{\rm ss}$. As mentioned in the main text, $U_{\rm ss}$  is proportional to $(\gamma_+ - \gamma_-)$, leading to $\Delta U = 0$  when $\gamma_+ = \gamma_-$ as showed in Fig.~\ref{fig:maxmixed_integrated}(a). At this point ($\gamma_+=\gamma_-$), the purity---not shown here---and entropy total variation from the initial to the SS show an extremum, as seen in Fig.~\ref{fig:maxmixed_integrated}(a). This extremum corresponds to a maximally mixed SS and, hence, both purity and entropy variations vanish in this case.  The same behavior holds for the total variation of `conventional' and EB heat in Fig.~\ref{fig:maxmixed_integrated}(a). Furthermore, in Fig.~\ref{fig:maxmixed_integrated}(b) we observe that the three approaches lead to the same evolution of irreversible entropy, as expected since the exchange of heat is identical in all of them.  More interestingly, we see a negative temperature along with  production of irreversible entropy during the transient, accompanied by a zero population inversion as the trajectory keeps in the maximally mixed state along the decay.


Starting from an initially pure ground state,  Fig.~\ref{fig:ground_time}(a) shows the evolution of the Bloch vector. In Fig.~\ref{fig:ground_time}(b) the evolution of the heat exchanged according to the three thermodynamics approaches is illustrated. The latter resembles to the evolution shown in Fig.~\ref{fig:thermo_t}(b) for the thermal state, namely: (i) the exchange of EB and conventional heat is similar and decrease monotonically, with oscillations when reaching the steady state, and (ii) the variation of internal energy (i.e. exchange of HB heat) drastically differs to the EB and conventional heat evolution, showing a maximum around $t \omega_0 \approx 2$. In contrast with the behaviour for the initial thermal state in Fig.~\ref{fig:thermo_t}(b), since the variation of EB heat at initial time does not coincide with the variation of internal energy in Fig.~\ref{fig:ground_time}(b), now the EB approach predict a production of work at initial time.  In Fig.\ref{fig:ground_gamma}(a) we observe that the variation of the thermodynamics quantities---from the initial to the steady state---as function of the relaxation rates exhibit a behavior similar to that of Fig.~\ref{fig:thermo}(a), which follows from an initial thermal state. However, we now observe that the net variation of entropy  is always positive, as expected for an initial pure state. Fig.~\ref{fig:ground_gamma}(a) shows again that, for maximally mixed SS (i.e. the vertical line at $\gamma_+ = \gamma_-$), the variations of purity and entropy as function of $\gamma_+$ exhibit an extremum while the variations of EB heat  present local peaks.  The evolution of the entropy and irreversible entropy for the different thermodynamics approaches is illustrated in Fig.~\ref{fig:ground_gamma}(b), with the evolution of the instantaneous inverse temperature of the system and the population inversion in the inset. We observe that the variation of entropy, as well as the inverse temperature, diverges at the initial time, as expected for a pure state. Additionally, we find again negative irreversible entropy production according to the HB approach.

\clearpage

\end{document}